# Cones, pringles, and grain boundary landscapes in graphene topology

Yuanyue Liu and Boris I. Yakobson

Department of Mechanical Engineering & Materials Science, Department of Chemistry, and the Smalley Institute for Nanoscale Science and Technology, Rice University, Houston, TX 77005, USA

A polycrystalline graphene consists of perfect domains tilted at angle α to each other and separated by the grain boundaries (GB). These nearly one-dimensional regions consist in turn of elementary topological defects, 5-pentagons and 7-heptagons, often paired up into 5-7 dislocations. Energy G(α) of GB computed for all range $0 \le \alpha \le \pi/3$, shows a slightly asymmetric behavior, reaching ~5 eV/nm in the middle, where the 5's and 7's qualitatively reorganize in transition from nearly armchair to zigzag interfaces. Analysis shows that 2-dimensional nature permits the off-plane relaxation, unavailable in 3-dimensional materials, qualitatively reducing the energy of defects on one hand while forming stable 3D-landsapes on the other. Interestingly, while the GB display small off-plane elevation, the random distributions of 5's and 7's create roughness which scales inversely with defect concentration, $h \sim n^{-1/2}$.

Graphite is a polycrystalline bulk material, whose three dimensions permit variety of grain orientations, grain boundaries (GB), several types of dislocations and point defects, all studied for decades.[1, 2] In contrast, reduced to the two dimensions sheet of graphene can not have same rich variety of imperfections and their types are more restricted. High resolution microscopy has made it possible to gain evidence of defects, even their temporal dynamics, in the context of carbon nanotubes[3, 4] and recently graphene[5-8]. The evidence of GB—the borders between the tilted perfect-crystal domains of single layer graphene—remains rather scarce[9-13]. Yet they must form when graphene islands nucleate at different sites of a substrate and the neighbour-islands are misoriented by some angles α. As such islands grow large and run onto each other, the GB form, Fig. 1a. What atomic organizations emerge when all covalent bonds are sealed at the GB at its lowest energy? What are the elementary constituent defects in the GB, what extra energies do they carry and how the total GB energy depends on the tilt angle? These are generic questions in any GB study, yet in the context of graphene not systematically addressed. The goal of this study is to quantify the regularities in the GB structures, generally disordered and complex at the atomics scale, their energy behaviors, and their manifestations in the non-planar distortions-warping.

Direct simulations by molecular dynamics (MD) are possible, by placing the misoriented graphene domains in contact within the plane and observing what morphologies emerge upon interface annealing. This remains of limited value, as the annealing is computationally costly while generated disordered structures offer limited insight, with the emerging 5-pentagonal and 7-heptagonal patterns are rather random (an excessive disorder problem which plagues the MD simulations of growth[14]). Here we choose different approach and analyze the structures and energies of different configurations, starting from most elementary "particles", the two types of Volterra disclinations[15, 16] in graphene: positive (5-pentagon) and negative (7-heptagon). From the analysis of their energies, significantly reduced by permitted off-plane relaxation, we note that their elastic energies diverge with the sample size much slower than in the case of bulk 3D-continuum. This suggests that a pair-dipole 5|7 should have converging energy, which can be evaluated, and serves a building block for the low-angle GB at $\alpha << 1$. Going beyond the low-angle case, we proceed to computing the energies of a whole range of GB, from the armchair interface ($\alpha_{AC} \approx 0$) towards zigzag interface ($\alpha_{ZZ} \approx \pi/3$). We find the GB energy function G($\alpha$) to follow an arch-curve, accompanied by interesting transition from one type of elementary dislocations 5|7 to another 5|6|7 through re-grouping of the 5's and 7's within the GB. Energy-reducing off-plane relaxation also manifests itself in possible formation of ridges and landscapes of substantial heights, observable with AFM[8] when the atomic resolution is not achievable to detect the GB.

To reveal the structures and evaluate the energies of the constituent defects and the GB, we use large scale energy minimization (preceded by finite temperature annealing, to ensure that we reach global minimum). By necessity, the system size (~$10^4$ atoms) makes the first-principle calculations impractical. Similarly, tight-binding approximation level is also insufficiently fast. Well developed classical interatomic potentials appear to be adequate for the task of finding general behaviors and regularities. Choice for hydrocarbons is Tersoff-Brenner type potential, more specifically AIRIBO,[17] as implemented in LAMMPS.[18] To extract the defect energy more accurately, we compute the total energies (either of large clusters, or with periodic boundary conditions, PBC) and subtract the total energies of perfect graphene systems of identical number of carbon atoms and identical perimeter of H-passivated edges.

It has been suggested in very early work[19] that the GB in graphene contain the pear-shaped polygons, later recognized as adjacent pentagons '5' and heptagons '7'.[20] These elementary defects can appear individually, or as 5|7-pairs, or as series of such pairs, in order to fit the inter-grain misorientation. It is important to appreciate that these seemingly "local" defects are qualitatively different from the true point defects (interstitials, vacancies, Stone-Wales transforms 5|7/7|5, etc.). All of the latter can be easily annealed by adding or removing an atom or two, or by rotating a bond back into its native position. In contrast, neither an isolated '5', or a '7', or their pair 5|7 cannot be corrected, annealed by *any* local reorganization of the lattice, either mass-conserving or not. Indeed, creating an isolated 5 in graphene requires a cutout of a whole "wedge" within an angle $\Delta = \pi/3 = 60°$ (positive disclination). To form a 7, one needs to seamlessly insert a similar wedge (negative disclination). In a 3D-bulk material such disclinations cause extreme deformations[16] and normally are energetically prohibitive. In a 2D-lattice, like graphene, they can relax through off-plane distortion, reducing the deformation energy to "affordable" levels discussed below (Fig. 1b).

In case of a 5-pentagon, simple geometrical analysis shows that a free lattice will form a cone (Fig. 1c), with axial angle $\delta = \sin^{-1}(1-\Delta/2\pi)$. It is useful to derive the energy of resulting elastic deformation (and compare it with the energy of computed atomistic structures), in order to highlight the difference from the regular point defects, having well defined formation energies. In case of a cone, the curvature at a distance r from the apex is $\kappa \sim 1/[r\cdot\tan(\delta)]$, and the elastic energy $\sim \kappa^2 \sim 1/r^2$ should be integrated over the entire cone, which yields $E = \pi D (\cos^2\delta/\sin\delta)\cdot\int \frac{dr}{r}$. While the natural lower-limit is the interatomic distance (bond length, *a*), at the upper limit the slowly decreasing integrand (~1/r) causes a divergence; the energy grows with the sample size as *ln*(R) and cannot be defined as intrinsic defect property. Instead, the defect energy depends explicitly on the cone radius, $E(R) = \pi D(\cos^2\delta/\sin\delta)(\,ln\ R - ln\ a)$. Fig. 1b shows that the data points from atomistic computations follow the $ln(R)$ dependence closely. Further, knowing the angle $\delta = 56.4°$, the slope of this line can be used to extract the flexural rigidity constant of graphene, D = 1.1 eV, close to reported.[21]

An isolated 7-heptagon—a similar disclination but of opposite sign—turns the geometry complicated: in contrast to the cone, here the axial symmetry is broken as the membrane bifurcates (yielding to the internal compression caused by extra material) into a shape of popular Pringles. From the 7-center of negative Gaussian curvature, the warped graphene canopy extends in all directions. Although exact equation for this extremal surface is not readily available, its self-similarity suggests that the curvature decreases as 1/r, and the total energy must grow with size as $ln(R)$, similar to a cone. The data points in Fig. 1b again follow the $ln(R)$ dependence very closely. Based on this agreement, one can conveniently write down the energies of these defects as $E = E_{core} + E_{elast}\ ln\ (R/a)$. Being assigned formally, $E_{core}$ can in principle be either positive or negative (if 1/r overestimates the strain near the 7). Important observation here is that individual 5 or 7 cause globally non-planar geometries, so called non-developable surfaces. This can be a likely a reason for variety of stable, non-fluctuative landscapes and wrinkles observed on graphene with atomic force microscope (AFM). Remarkably, the cones of graphene has been synthesized in all five varieties[22] corresponding to the 1, 2, 3, 4, or 5 pentagons at the apex (and the angles $\delta$ = 56.4°, 42°, 30°, 19.5°, 9.6°, respectively; 6 pentagons correspond to $\delta = 0$ that is parallel walls of a nanotube). In contrast, we are not aware of any observations of graphene pringles, although they have comparable or lower formation energies, Fig. 1b.

Since the energies of the 5's and of 7's are quite large, caused by the delocalized lattice strain, their pairing up into dipoles is energetically favorable. By analogy to electrostatics, such dipole of positive $+\Delta$ and negative $-\Delta$ disclinations is expected to produce a strain field $\sim 1/r^2$ at the distance r [a derivative of the field from the monopole, $\partial/\partial r(1/r)$], so that the elastic energy density of the surrounding lattice falls as $1/r^4$, and thus its integral value should converge. Thus, one should be able to define the energy of a 5|7, in contrast to the 3D crystals where dislocation energy diverges logarithmically. Having a 5 and a 7 nearby in a lattice creates a dislocation of arbitrary Burgers vector.[23] The simplest are the well known 5|7 of Burgers vector **b** = (1,0) in notations of nanotubes (or ⟨2,-1,-1,0⟩ in crystallographic notations for graphite), or a 5|6|7 of Burgers vector **b** = (1,1).[20] A full relaxation of a series of graphene islands with single 5|7 yields the values E(R) varying insignificantly, much slower than for the bare disclinations, Fig. 1b. Computational limitations prevent one from reaching a clear asymptotic value. It can be evaluated from consideration of dislocation walls, which we are turning to next.

A GB separates the domains tilted to some angle $\alpha$. Typically it bisects the tilt at $\alpha/2$ angle relative to the crystal planes of the grains. Generally, it may deviate by some angle $\beta$ from the bisector, dividing the tilt angle into $\alpha/2 \pm \beta$. The GB must contain imperfections to accommodate for the tilt. Macroscopically, considering a closure failure around the contour shows that the Burgers vector density per unit length is $d\mathbf{B}/ds = 2\sin(\alpha/2)(\mathbf{n}\times\boldsymbol{\alpha})$, where $\mathbf{n}$ is a unit vector along the boundary and $\boldsymbol{\alpha}$ is a tilt vector.[15] Beyond this macroscopic Frank equation, to explore the details of the GB structure and energy, we perform the systematic atomistic calculations. Before considering the role of the tilt angle $\alpha$, we perturb the boundary by gliding (via the Stone-Wales rotations[20, 24]) the constituent 5|7-cores, Fig. 2a-c. The plot in Fig. 2d shows increase in energy. This agrees with the well known preferred vertical alignment of two identical edge dislocations, ⊥ (in contrast to a pair of opposite dislocations: ⊥ and its inversion twin, which form a stable dipole aligning at 45°). The GB nearly bisecting the tilt angle must be therefore main choice of detailed energy analysis. Nevertheless, low mobility of constituent dislocations suggests that the no-bisector GB, if formed in the course of growth, can be kinetically stabilized and encountered in observations as well.

One example of interest is a possible 30°-tilt interface formed by a zigzag (ZZ) domain edge on one side and an armchair-oriented (AC) domain on the other, as shown in Fig. 2e. Formally, they can be matched by a series of close packed and slanted 5|7's, as shown. (Structure closely resembles the reconstructed ZZ edge, reczag[25]). This appears as topologically satisfactory solution, but the period of the zigzag ($2\sqrt{3}a$) exceeds the period of armchair side domain ($3a$) by 15.5%. Such mismatch would cause a high energy distortion. In order to form a well-matched GB, a certain number of atomic rows (about every $8^{th}$ and occasionally $7^{th}$) should be removed on the ZZ side, which is equivalent to insertion of sparsely spaced 5|7 cores. Upon full relaxation, one obtains a GB without remote stress. In the sites of the extra 5|7 insertions, it displays a peculiar "fly-head" 7/5\7 structure, where the standard pentagon-heptagon dislocation cores appear flip-altering their orientation from 5/7 to 7\5. Since the AC|ZZ mismatch is an irrational number, the locations of the fly-head pattern cannot be periodic. This one example illustrates the rich realm of possibilities in the prime units (5-pentagon and 7-heptagon) organization of low energy GB. As expected, the computed energy of this interface is higher than the bisector types for the ~30°-tilt.

Now we consider the subset of GB which are the bisectors, as a most realistic choice, and turn to the question of how they change with the tilt angle. Since the direct computations of numerous possibilities are extensive, a preliminary analysis gives some guidance. For low angle GB, the generic structures are well discussed:[15, 16] it is a series of edge dislocations, of Burgers vector b (in case of 5|7, it is one lattice parameter, $b = \sqrt{3}a$) and spaced by a distance $b/[2\sin(\alpha/2)] \sim b/\alpha$. If the energy E of individual dislocation is defined, then the GB energy is roughly proportional to the density of these defects, $G \sim \alpha\cdot(dG/d\alpha)_0 \sim \alpha\cdot E/b$. Increasing the tilt $\alpha$ makes the dependence nonlinear, but the opposite limit becomes simple again. For graphene, at $\alpha = \pi/3$ the perfect lattice is fully restored, while in its vicinity the dislocations sparsely placed along the GB give rise to its energy, $G \sim -(\pi/3 - \theta)\cdot (dG/d\alpha)_{\pi/3}$. It is easy to see that in this "near ZZ" limit the elementary dislocation is different, a 5|6|7 of larger Burgers vector $b` = 3a$, larger energy and consequently different slope. Further, the overall functional behavior $G(\alpha)$ should be periodic, sought as a sum of a few Fourier overtones, with a leading term $\sim \sin(3\alpha)$.

Fig. 3 summarizes the results of energy computations for a number of constructed GB from nearly AC contact (small $\alpha$) to the nearly ZZ interface ($\alpha \approx \pi/3$). Fig. 3b-f show the GB structures, placed near their respective tilt angles. The overall energy $G(\alpha)$ behavior is rather close to sine-function, although non-equivalence of the left- and right-limit structures (AC and ZZ interfaces) causes some asymmetry; the maximum is not necessarily in the middle ($\alpha = \pi/6$) yet it appears rather close. Overall range of energies is up to 4.5 eV/nm, much lower than the for the bare graphene edges (~10-13 eV/nm, depending on the type[25]).

The energy-arch is interesting to follow from left to right, to understand the logic of GB structure changes as the tilt angle changes. At small values it is simply a series of separate 5|7, Fig. 3b. As the angle increases, they get closer to each other and become crowded. Eventually, we reach the highest density in a sequence (**5|7**)6(**5|7**)6(**5|7**)6(**5|7**)6…; here the 5|7's are separated by single hexagons only, Fig. 3c. Further tilt increase causes peculiar regrouping, when pentagons and heptagons abandon the original partners (by insertion of one 6-hexagon) and pair up with the ones on the opposite side, as **5**)6(**7**|**6**|**5**)6(**7**|**6**|**5**)6(**7**|**6**|**5**)6(**7**|**6**…. After that, the energy descent corresponds simply to increasing spacing between the new elementary dislocation cores 5|6|7 which eventually leads to the ZZ interface. Fig. 3b-f show the important intermediate GB structures, as well as the simple cases of the low-tilt boundaries near AC ($\alpha = 0$) and ZZ ($\alpha = \pi/3$) edges, where the two grains merge perfectly, $G(0) = G(\pi/3) = 0$. While the full fit in the Fig. 3 was done with three harmonics, a behavior is roughly captured by approximation $G(\alpha) = 5\cdot\sin(3\alpha)$, eV/nm (the next two coefficients are 0.1 and 0.3). Besides the overall energy behavior, one can evaluate the energy of a single 5/7 from the low-angle limit, in which case the leftmost point in the plot gives $E_{5/7} \approx 5$ eV. This value is marked in the Fig. 1b (gray horizontal line) and is the asymptotic value for the single-5/7 data, apparently reaching this limit from below.

In our computations we rely on PBC along the GB direction, but consider limited width in perpendicular x-direction (after checking the results insensitivity to further increase of this width). Rapid decrease with the distance x from the boundary is illustrated by the gray-level local strain energy representation (per atom) in Fig. 3b. This near-field ($x < b/\alpha$) analysis complements the known analytical result for the strain energy, $\sim x^2\alpha^4\, e^{-4\pi x\alpha/b}$ at $x > b/\alpha$.[15]

Noting that in 3D-crystals the elastic energy is usually lowered by splitting the dislocations $(1,1) \rightarrow (1,0) + (0,1)$ to smaller Burgers vectors,[15] we consider a GB in Fig. 3f as alternative to Fig. 3e. In contrast to 3D-continuum, in a freestanding 2D-lattice such a split is unfavorable and the energy is ~ 1 eV/nm lower for the 5|6|7 cores.

This example of the larger Burgers vector being energetically preferred (Fig. 3e-f) as well as convergence of the 5|7 energy in a free graphene is due to additional freedom to warp off-plane. As a result, the initially 2D-flat sheet forms rather pronounced 3D-lanscapes. Extreme manifestations of this are of course the cone and pringle, which cannot be developed onto a plane at all. Their pairs can, yet both small 5|7 and especially large Burgers vector 5|66666|7 dislocations cause great distortions, as computed structure in Fig. 4a shows. The roughness of emerging stable landscape can be characterized by the heights, $h_{5\text{-}7}$. Although obviously random, the typical height can be estimated from our knowledge of the topology-induced conical shape (with $\cos\delta \sim 0.5$) and the peak-valley distances $\sim b_{5\text{-}7}$ (that is 5-7 distance, proportional to the Burgers vector associated with a 5-7 dipole): $h_{5\text{-}7} \sim b_{5\text{-}7}$. Remarkably, we see that the randomly scattered disclinations at concentration n of pentagons and heptagons cause the roughness inversely proportional to the defect concentration, $h_{5\text{-}7} \sim n^{-1/2}$. In this context an interesting extreme case of highest defect concentration ($n_{5\text{-}7} \rightarrow \infty$) is the 5's and 7's closely packed into so called pentaheptite crystal of planar geometry ($h_{5\text{-}7} \rightarrow 0$).[26]

If organized in linear motifs of grain boundaries, they form linear ridges, as one in Fig. 4b. When grown or placed on a substrate, topologically-induced graphene landscape is partially flattened by the "gravity" of van der Waals attraction (energy V per area). To estimate the resulting elevation, we note that there is only one other essential parameters, the flexural rigidity D. Dimensionality consideration yields the elevation $(h - h_{flat}) \sim (D/V)^{1/2}$. Two ridge-profiles in Fig. 4c, computed at two different strengths of attraction (contact surface energy, V) show good agreement with this dependence.

One can easily imagine that the lattice distortions near the GB do change the electronic structure in its vicinity, and therefore will cause scattering of electrons, affecting the transport phenomena both across and along the GB direction. These aspects, although both interesting and potentially important, requires further study beyond the scope of present report.

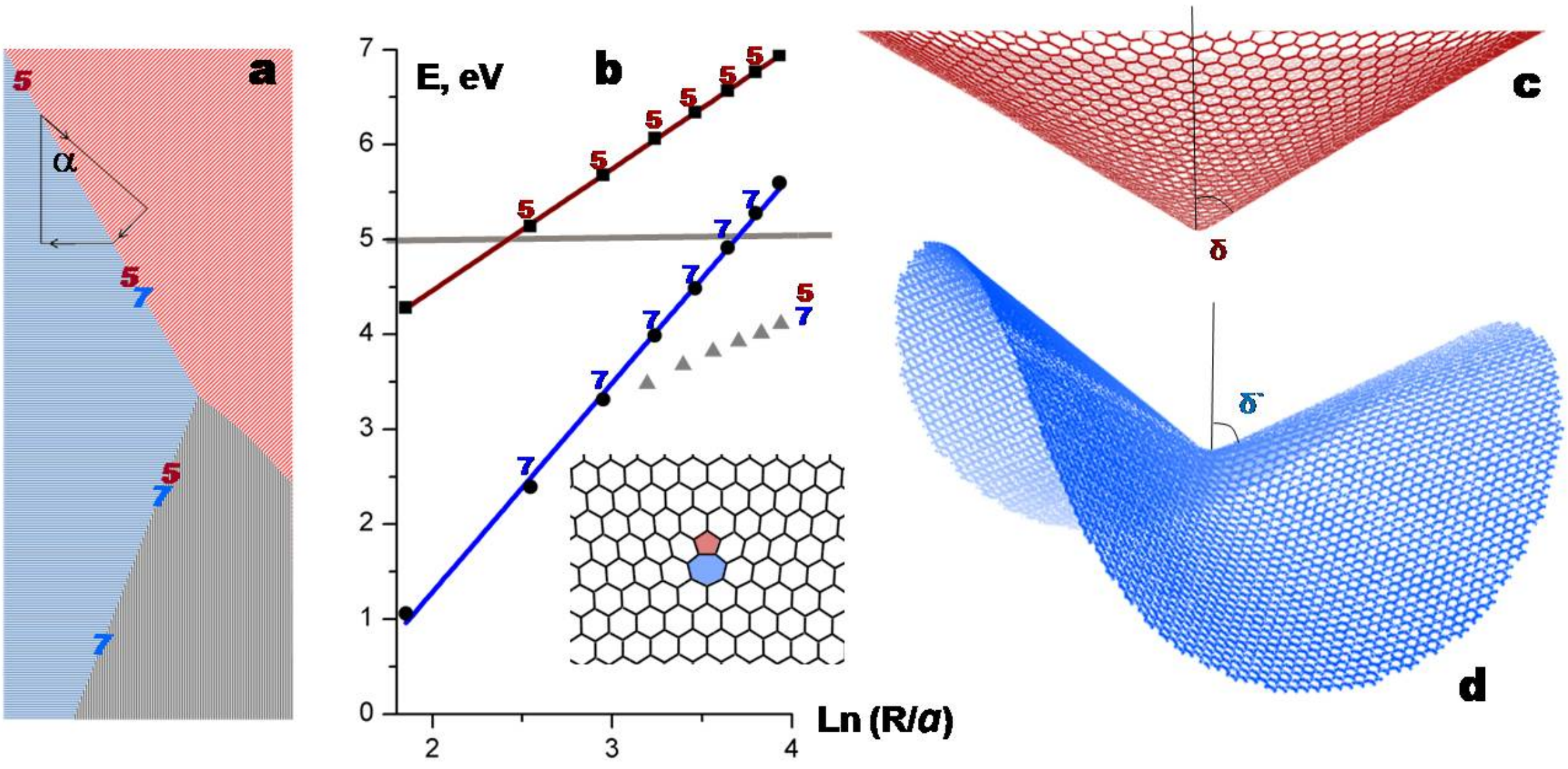

**Figure 1.** (a) 2-dimensional domains misoriented by the tilt angle α are separated by the grain boundaries (GB), made up of 5-pantagons and 7-heptagons. (b) Defect energies computed for an isolated 5, a 7, and a 5|7-dislocation, as a function of size R of the lattice cluster. (c) Fully relaxed lattice containing a 5 becomes a non-planar cone. (d) Graphene lattice containing a single 7 warps into a shape of nano-Pringle, with δ` = 63°.

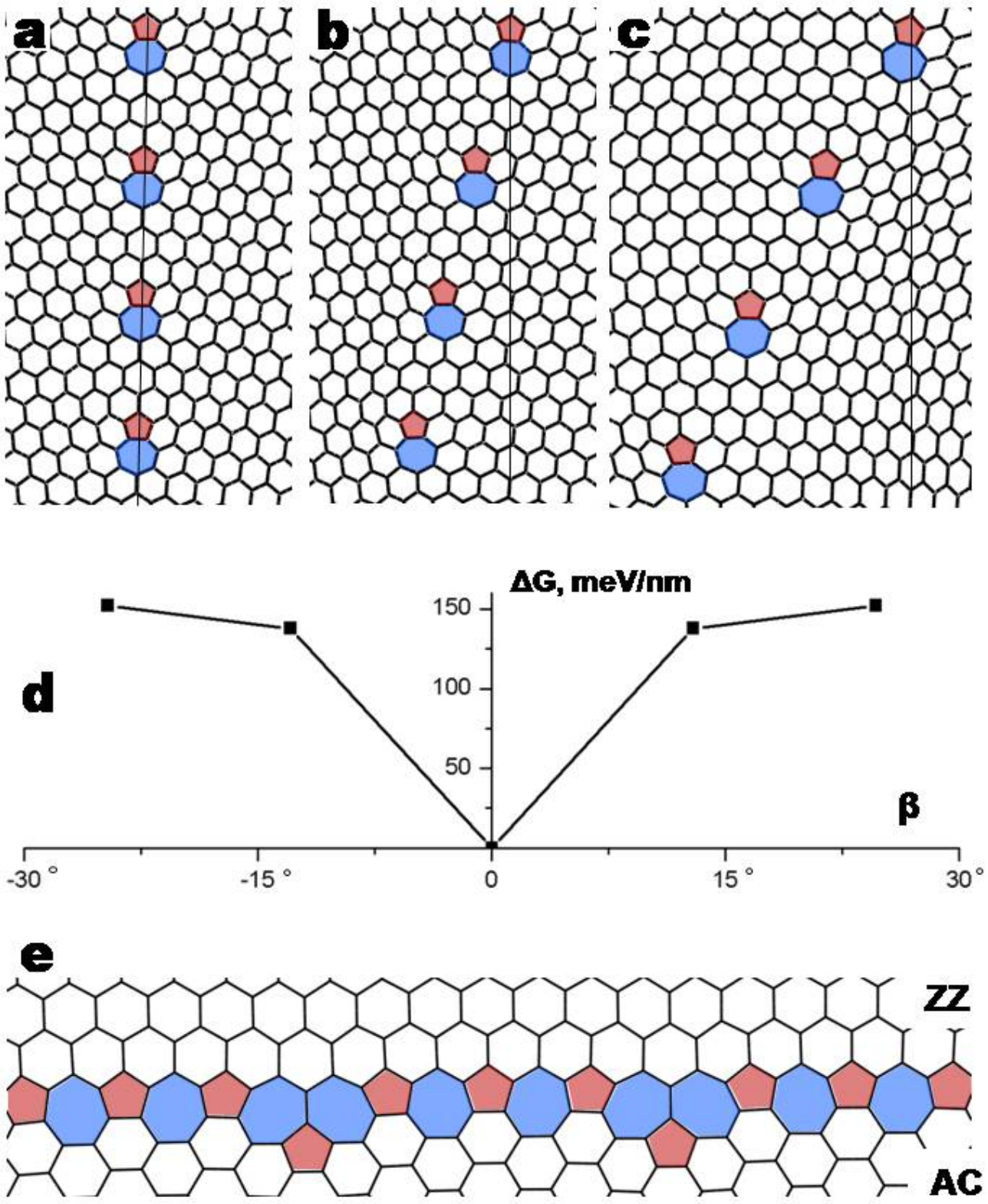

**Figure 2.** Simple GB structure in its generic bisector position (a), slanted by β = ±12° (b) and β = ±24° (c), and their relative energies (d). In (e) an interface joining a ZZ edge and an AC edge between the domains tilted to α = 30°; note the "fly-head" pattern where the atomic rows removed to reduce the interface mismatch strain.

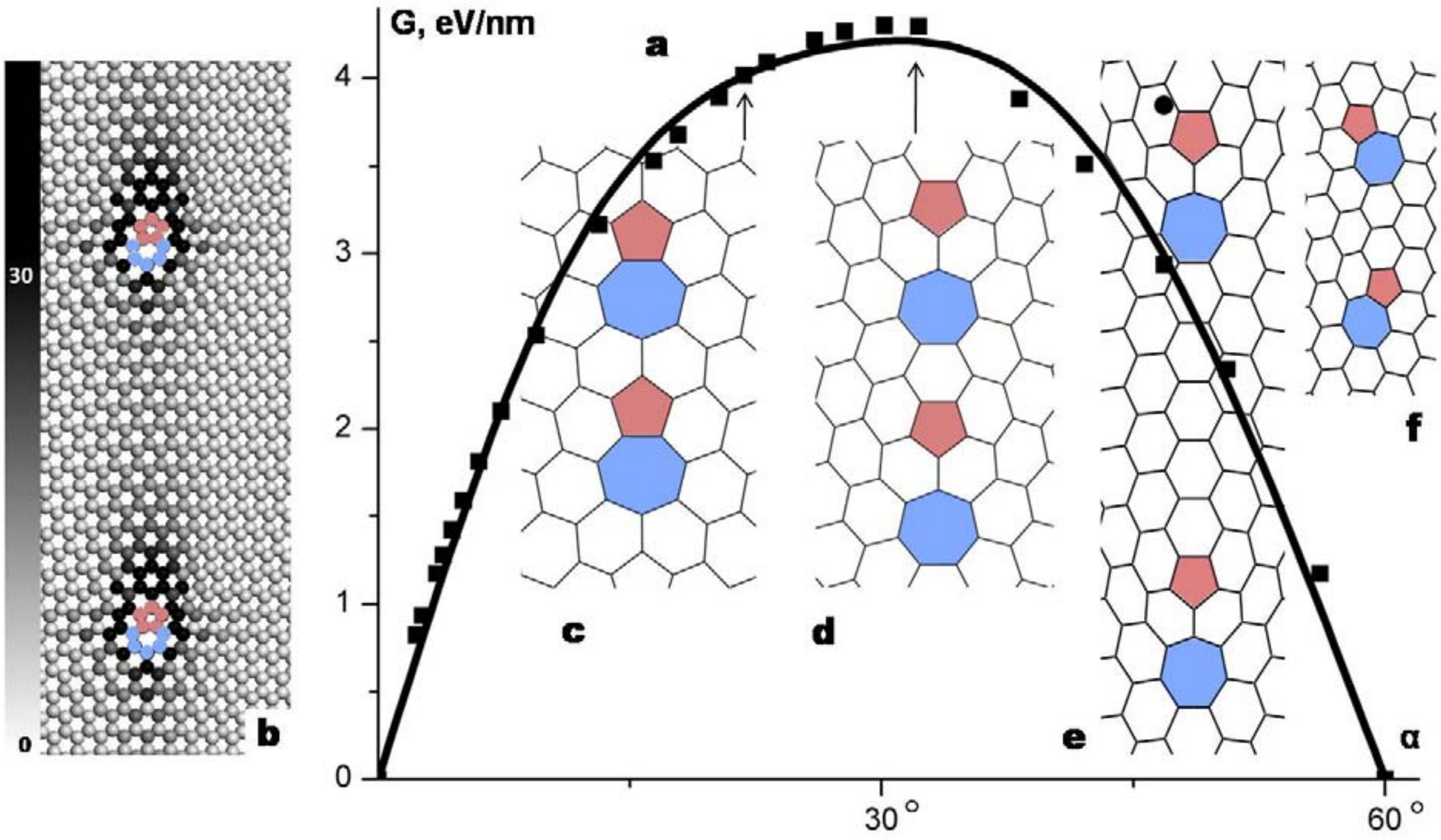

**Figure 3.** (a) Grain boundary energy G(α) as a function of tilt angle, based on 26 computed data points (solid squares) and fitted by the three-terms of Fourier series (thick line). (b) Nearly-AC interface low-angle GB (α = 3.5°) with gray-level coded strain energy per atom, shown in meV. (c) Maximum 5|7 density GB, α = 21.8° (d) Maximum 5|6|7 density GB, α = 32.2°. (e) Nearly-ZZ interface low-angle GB (α - 60° = 13.2°) comprised of 5|6|7 and (f) its alternative split into slanted 5|7's, higher in energy (solid circle).

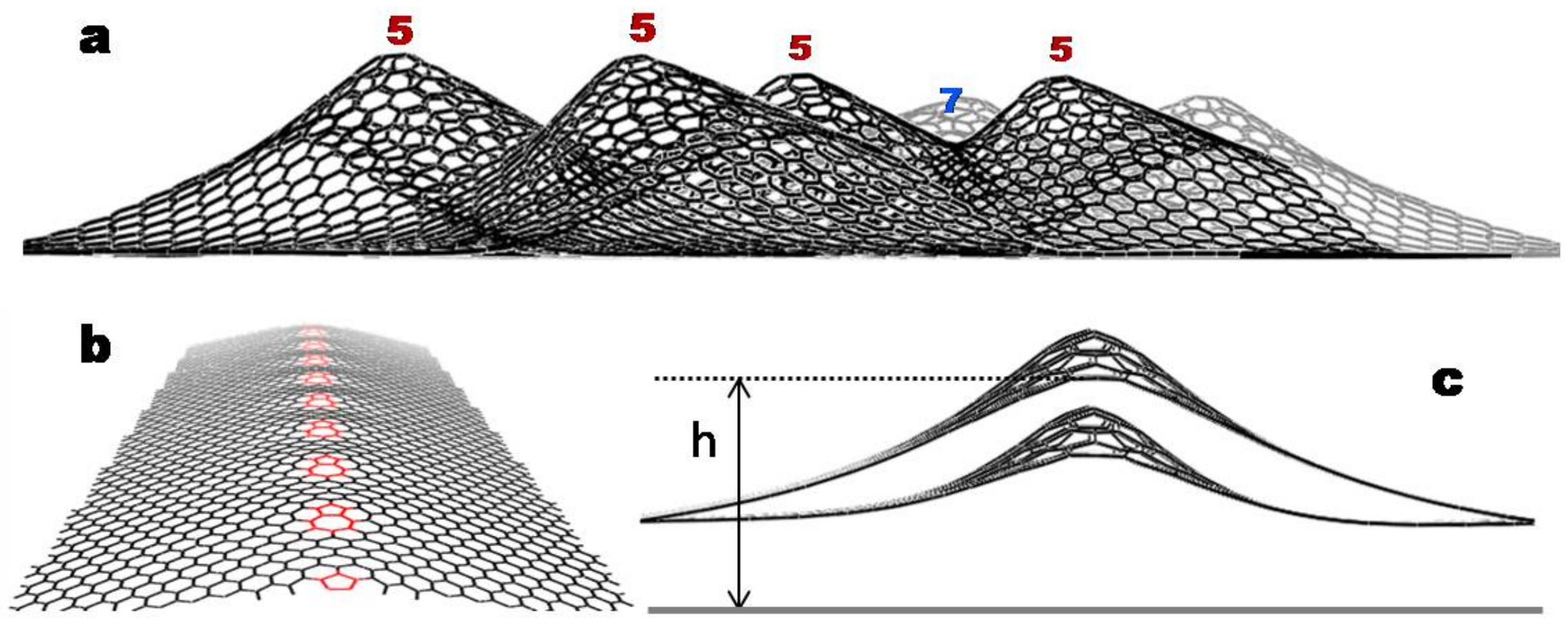

**Figure 4.** (a) A computed (full energy relaxation) landscape due to presence of scattered 5 and 7 defects in a perfect graphene lattice, shows elevation roughly equal to the distances between the 5's and 7's. (b) A regular GB from 5|7 dislocations forms a ridge shape. (c) Flattening effect of the van der Waals attraction to the substrate, as computed for the two values 4V and V, when the elevation $h - h_{flat}$ approximately doubles.